\documentclass[copyright]{eptcs}
\usepackage[utf8]{inputenc}
\usepackage{url}

\title{Sending Money Like Sending E-mails: Cryptoaddresses,\\The Universal Decentralised Identities}

\author{Michal Zima
\institute{Faculty of Informatics,
Masaryk University,\\
Brno, Czech Republic}\\
\email{xzima1@fi.muni.cz}}

\def\titlerunning{Cryptoaddresses}
\def\authorrunning{M. Zima}

\begin{document}
\maketitle

\begin{abstract}
Sending money in cryptocurrencies is majorly based on public keys or their hashed forms---``addresses.'' These long random-looking strings are user unfriendly for transferring by other means than via copy-and-paste or QR codes. Replacing such strings with identifiers chosen by users themselves would significantly improve usability of cryptocurrencies. Such identifiers could be memorable, easier to write on paper or to dictate over phone. Main challenge lies in designing a practically usable decentralised system for providing these identifiers. Former solutions have been built as centralised systems or come with nonnegligible limitations. Our solution is reminiscent of a prevalent e-mail system, which is an already user friendly and desirably decentralised system. It is shown that our approach is directly applicable also to other systems that use long cryptographic identifiers.
\end{abstract}

\section{Introduction}
Nowadays, the most common way for sending money in cryptocurrencies is to copy-and-paste payee's address---usually a string of over 30 random characters long hash of their public key---or to scan its QR code representation. For instance, in Bitcoin such address looks like this: 1\-N\-S\-1\-7\-i\-a\-g\-9\-j\-J\-g\-T\-H\-D\-1\-V\-X\-j\-v\-L\-C\-E\-n\-Z\-u\-Q\-3\-r\-J\-D\-E\-9\-L. However, it is far from being user friendly in terms of manual transfer---on paper, over phone or in one's memory.

A primary requirement for a system providing mapping of user friendly identities to the original user unfriendly strings is to be decentralised. Centralised systems usually already provide user friendly identities, while decentralised trustless systems often do not. It is crucial for the new system to keep up with the systems it provides identities for, i.\,e., not to be centralised (and potentially a single point of failure) in a decentralised environment.

Our system of cryptoaddresses, presented in this paper, introduces e-mail-like identifiers, leveraging the existing DNS system. Usage of DNS provides straightforward decentralisation on the level of ``domain namespaces,'' designating a server of user's own choice for each of their domain names. Servers provide resolution of cryptoaddresses to the original cryptographic identifiers. The system is secured both on the level of DNS and communication with servers. The scope of applicability is not limited to cryptocurrencies---the system is universal with variety of possible applications.

The remainder of this paper is organised as follows. Section \ref{sec:related_work} presents previous work related to the problem. The overall design of the system is discussed in section \ref{sec:design}, including definitions of used identifiers. Section \ref{sec:dns} elaborates on details of utilisation of the DNS system, followed by a design of our protocol SCAP in section \ref{sec:scap}. Security considerations are presented in section \ref{sec:security_considerations}. We discuss possible future work in section \ref{sec:future_work} and provide final conclusions in section \ref{sec:conclusion}.

\section{Related Work}
\label{sec:related_work}
Endeavours to make complex addresses more usable began within Bitcoin by a concept of ``first bits.'' First bits make use of implicit information contained in cryptocurrency's database of transactions---blockchain. Using first bits, an address can be identified by a short prefix of only several characters. This prefix is expanded to a full address by looking up in the blockchain the first address matching the prefix \cite{firstbits}. However, first bits have also several disadvantages. Most notably, prefix lookup and obtaining a new prefix are time-expensive operations. Besides, it is not guaranteed to be good-looking, the address has to be already used, i.\,e., to have some money sent to it (this fact alone goes against a common practice of not reusing addresses in cryptocurrencies), and the prefix cannot be revoked or reassigned.

Later on, many centralised systems arose, e.\,g., Keybase \cite{coindesk_keybase_2014} or Gravatar \cite{gravatar}. Nonetheless, none of them has ever been adopted by the cryptocurrency community.

In 2014, community of a cryptocurrency called Monero created a project OpenAlias \cite{openalias}. OpenAlias enables translation of domain names to addresses of various cryptocurrencies. This is achieved by specially crafted TXT records in DNS zones of respective domain names. While the record can be secured with DNSSEC against tampering, similarly to first bits, revocation and replacement of the cryptocurrency address included in the record is a subject to expiration period (TTL) of DNS caches, therefore a change in the record may be visible after delay.

All current approaches share a drawback in their design: they are one-to-one mappings. It is not possible to dispense a different cryptocurrency address to every user. Instead, all obtain the same one. This is a nonnegligible flaw in terms of privacy preservation. Not only is publicly known the connection between the identity and the address, but the address itself gets reused, making linkability of different payments to a single entity trivial.

\section{Design of Cryptoaddresses}
\label{sec:design}
Our scope is a decentralised system for providing users with arbitrary identities. Primary goal of this system is to translate these identities to addresses in any cryptocurrency. We call this system \emph{cryptoaddresses}.

Systems based on a shared database, e.\,g., first bits, retain impractical properties related to the database use and maintenance. In contrast, systems built upon existing DNS infrastructure can leverage its inherent distribution of responsibility. As with e-mail, every administrator of a domain name designates servers responsible for receiving e-mails for this particular domain name. Thence, cryptoaddresses are made decentralised by utilising DNS, and further employing DNSSEC to secure information stored in and retrieved from the DNS.

\subsection{Format of Cryptoaddresses}
\label{sec:format}
A cryptoaddress (also called a \emph{Crypto ID}, abbreviated as \emph{CID}) is formed of a \emph{local part}, an \emph{@ symbol} and a \emph{domain part}: local-part@domain-part. A cryptoaddress intentionally resembles an e-mail address, because of familiarity of these addresses to users---the format is nowadays easily recognisable and is known to form an address. The same format is also used in XMPP to form an \emph{XMPP address} (or a \emph{Jabber ID}, \emph{JID}) \cite{xmpp-address}.

Since the local part may also contain @ symbols, always the last one is used as a separator of the local and domain part. Format of the local part is not restricted and may be formed of any valid UTF-8 string. Length of the local part may be up to 1023 bytes; zero length is allowed. This length limit is high enough for any practical use case and at the same time gives a concrete upper bound useful for implementers.

The domain part must conform to the format of a fully-qualified domain name (FQDN) \cite{fqdn}, without an ending dot. Internationalised domain names are allowed---conversion to standard ASCII format of domain names is done in a way specified by the IDNA standard \cite{idna2008}.

\subsection{Identifiers of Addressable Services}
Single cryptoaddress may be used as identity for various systems and services at the same time. To internally differentiate one from another, each such addressable service must be uniquely identified. This is especially of a concern within cryptocurrencies as there are cryptocurrencies with ambiguous names and/or codes (or even multiple codes). To provide a guidance for such situations, we propose usage of a hash of their genesis block (i.\,e., the very first block in the block chain) in hexadecimal encoding, which provides unique identification not only among other cryptocurrencies, but also among other systems. Examples for few cryptocurrencies are shown in table \ref{tab:cc_ids}.

\setlength{\tabcolsep}{4pt}
\begin{table}
\begin{center}
% The final sentence of a table caption should end without a period
\caption{Examples of service identifiers for several cryptocurrencies}
\label{tab:cc_ids}
\begin{tabular}{ll}
\hline
\noalign{\smallskip}
Bitcoin  & 000000000019d6689c085ae165831e934ff763ae46a2a6c172b3f1b60a8ce26f\\
Litecoin & 12a765e31ffd4059bada1e25190f6e98c99d9714d334efa41a195a7e7e04bfe2\\
Dogecoin & 1a91e3dace36e2be3bf030a65679fe821aa1d6ef92e7c9902eb318182c355691\\
\noalign{\smallskip}
\hline
\end{tabular}
\end{center}
\end{table}
\setlength{\tabcolsep}{1.4pt}

For other kinds of services, e.\,g., a communication platform Bitmessage\footnote{\url{https://www.bitmessage.org/wiki/Main_Page}}, a single identifier needs to be established by them. We do not suggest any concrete identifier for individual services. Nevertheless, it may be suggested to use lowercase ASCII characters for these identifiers, mainly for their ease of use and recognition.

Note that these service identifiers are important for flawless internal functioning of the system while users usually do not encounter them. For details on their usage in the protocol see section \ref{sec:scap:protocol}.

\subsection{System Components}
\label{sec:components}
The system is designed of several components: a client, a cryptoaddress, a SRV DNS record, a cryptoaddress server and a protocol for communication with the server. The cryptoaddress server is specific for the domain given in the cryptoaddress. For a single cryptoaddress there might be more than one such server---for load balancing or high availability purposes, as detailed in section \ref{sec:dns}.

A client communicates with a cryptoaddress server via a Simple Cryptoaddress Protocol, described in section \ref{sec:scap}. The communication is encrypted with a shared key derived from server's public key and client's private key using ECDH algorithm \cite{ecdh}.

\section{DNS Records}
\label{sec:dns}
An intermediate step in the translation of a cryptoaddress to target data is obtaining an address of the translating server. This address is provided by DNS infrastructure. Out of available DNS record types we choose SRV record. It is a record type dedicated for service discovery, i.\,e., for this kind of information, and provides great flexibility for designated server's FQDN, L4 protocol, port number, priority and weight within a set of servers designated for the same application protocol \cite{dns-srv}.

In DNS zone format the record has the following form:

\begin{verbatim}
_service._proto.name. TTL class SRV priority weight port target.
\end{verbatim}

We propose the service identifier \texttt{scap} and a TCP port 4332. Nonetheless, a server administrator may choose to use arbitrary port number since clients retrieve information about it in the SRV record.

\subsection{FQDN Formation}
\label{sec:dns:fqdn_formation}
Traditional approaches for establishing a secure connection to a server include X.509 certificates and trust-on-first-use (TOFU) trust model with server's public key provided by the server itself. Nonetheless, to simplify both the secure key distribution and the communication protocol, we leverage an idea originally found in DNSCurve \cite{dnscurve}. Instead of sending the server's public key during the initial protocol handshake, we encode it in the server's name. Therefore, a client obtains the key from server's FQDN which is included in the SRV record and even the very first message sent to the server can be hence encrypted. Security achieved with this approach is higher than with TOFU model due to the secure key distribution. In this section we describe how the key encoding into FQDN is done.

We place the key preceded by a version prefix into a dedicated FQDN label. By ``FQDN label'' we understand a part of FQDN delimited by dots from other FQDN labels. For encoding of the version prefix and public key we use the same base-32 encoding as DNSCurve. The specification of used base-32 encoding is provided within the DNSCurve specification in \cite{dnscurve}. Cryptoaddresses are likely to be used together with DNSCurve, therefore it is advantageous to use single encoding algorithm. Besides, a standard base32 encoding algorithms are not suitable for use in FQDN due to their usage of padding character \textit{=} (equals sign) \cite{rfc-base32}.

Version prefix is a four-character string \texttt{1000} which is formed by a base-32 encoded 2-byte little endian number \texttt{1}. Encoding of 2 bytes always results in 4 characters, therefore this FQDN label can be always safely split to a version prefix and a key before performing an actual base-32 decoding. If some further version of cryptoaddresses uses more labels in a FQDN, version prefix must be always present in the leftmost label. Version does not only determine the actual FQDN creation, it also indicates version of the protocol to be used for communication with the server. This way there is no need to indicate version in the messages sent between a client and a server and compatibility can be resolved before initiation of the communication.

After the version prefix a base-32 encoded public key follows. As a cryptographic scheme for the key we use Curve25519, a scheme based on elliptic curves. Its choice is based on its well-founded choice of elliptic curve and constants, availability of high-grade implementations and small size of keys, while providing reasonable security \cite{curve25519}. Although small size of public keys is not necessary, it is still beneficial as it allows keys to be directly used. Still, usage of elliptic curve cryptography makes it possible to leverage ECDH algorithm for establishment of a shared key without any additional communication \cite{ecdh}.

\subsection{Deployment Considerations}
When a cryptoaddress server serves many domains or domains whose DNS zones are out of control of the cryptoaddress server operator, it may be more convenient to use an alias in the domains' SRV records. An alias can be, for instance, \emph{scap1.example.com} pointing via a CNAME DNS record to the canonical name as described above. This provision also makes it easier to rotate keys used by cryptoaddress servers (e.\,g., when corresponding private keys are exposed during a security breach), as it requires change to only one DNS record per cryptoaddress server.

It is possible to operate several cryptoaddress servers, e.\,g., to enhance availability of cryptoaddresses within a domain. However, their operator must ensure that they provide equivalent information. In a case one server is missing information about a pair of a certain cryptoaddress and a service, client processes this information as authoritative and does not repeat their request to other servers.

An example of SRV records for a setup of two load-balanced servers and one backup server for a domain example.org follows (each record is broken into 2 lines by ``\textbackslash'' in order to fit into this limited space):

\begin{verbatim}
_scap._tcp.example.org. 86400 IN SRV 10 65 4332 \
    1000vs2nh9b3gz04db4rgpjmzv2cwlnpvh3qzn6xljwyxmnp57j8h0d.example.org.
_scap._tcp.example.org. 86400 IN SRV 10 35 4332 \
    100027q245f6cglhdjyy91vk5btyszk6g5fnhz7mvsc6mtfjh2q0c14.example.org.
_scap._tcp.example.org. 86400 IN SRV 30  0 4332 \
    10009ydzvtccqmbzw6q0zlgumtr227g0kwb2zk8h5rv7yruj7gg6zh3.example.org.
\end{verbatim}

The first two servers have priority 10 (which is higher than priority 30) with weights 65 and 35, meaning that requests to these two servers will be statistically divided by ratio 65:35. If both of them are unavailable to clients, a server with the next lowest priority is selected, i.\,e., the server with priority 30 in this case. All three servers use TCP port 4332. The last part of each record (all starting with \textit{1000...}) is the actual address of the cryptoaddress server with encoded version and public key.

\section{Simple Cryptoaddress Protocol}
\label{sec:scap}
Simple Cryptoaddress Protocol (abbreviated \emph{SCAP}) is designed for communication of clients with cryptoaddress servers. It is a synchronous protocol minimising the number of needed round trips (i.\,e., request-reply pairs), while maintaining its universality and extensibility. Its security design is inspired by DNSCurve extension of DNS \cite{dnscurve}---it shares with DNSCurve the scheme for keys and their exchange, nonce extension technique, or use of cryptographic boxes. Nonetheless, the binary format differs and different is also our choice of algorithm for cryptographic boxes (for authenticated encryption).

Distribution of servers' public keys is carried out by embedding them in FQDN as described in section \ref{sec:dns:fqdn_formation}, therefore every client has a public key of the server they connects to. A client provides the server with their public key in the first message they sends to it. In order to prevent connection hijacking, the server remembers client's public key for the duration of the session and the client never resends their public key within the session.

For cryptographic boxes we use IETF standard ChaCha20-Poly1305 authenticated encryption algorithm \cite{ietf-cryptobox}.

\subsection{General Message Format}
Format of messages slightly differs depending on whether a message is sent by a client or a server and whether it is client's first message in the session or not.

\subsubsection{First Client's Message}
\begin{itemize}
\item 32 bytes: client's public key.
\item 6 bytes: client's nonce choice.
\item Cryptographic box with the following content: \texttt{H}
\end{itemize}

\subsubsection{Follow-up Client's Message}
\begin{itemize}
\item 6 bytes: client's nonce choice.
\item 6 bytes: server's nonce extension.
\item Cryptographic box with the actual message content.
\end{itemize}

\subsubsection{Server's Message}
\begin{itemize}
\item 6 bytes: client's nonce.
\item 6 bytes: server's nonce extension.
\item Cryptographic box with the actual message content.
\end{itemize}

\subsubsection{Nonces}
A nonce in this context is a 96-bit number that is unique for each message within a single shared key (otherwise the shared key would be exposed). Half of the nonce is chosen by the client and half by the server. While the basic principle for nonce creation is shared with DNSCurve, details differ due to session-oriented nature of the communication in SCAP. Mainly, only the first client's message uses zero server's nonce extension for the cryptographic box---any follow-up client's message repeats the extension from the last server's message. Similarly, server always repeats client's chosen part of the nonce from their last message. Since neither client nor server repeat their part of the nonce, each message is guaranteed to have a unique nonce.

\subsection{Protocol Messages}
\label{sec:scap:protocol}
The protocol is client-driven---the server only sends responses to client's requests. Both requests and responses use a format based on the netstring encoding, as defined in \cite{qmtp}. In brief, a netstring has a format \texttt{[len]:[string],}, where \texttt{[len]} is a decimal representation of length of \texttt{[string]} in bytes. Netstrings are usually catenated, but are also allowed to be nested.

The first byte of a message determines a type of the message. We specify the following types of client's messages:

\begin{description}
\item [H] A hello message, sent right after opening the connection. The message content consists only of this byte. The primary purpose of this message is to provide the server with the client's public key.
\item [Q] A query request. It consists of two netstrings (nested inside of the main message netstring): a cryptoaddress to be resolved and a service identifier.
\item [E] This message type is reserved for possible use by future protocol extensions. Nonetheless, its exclusive use by extensions is not strictly required---an extension may also define new message types.
\item [X] A type reserved for proprietary/non-standard protocol extensions. The message consists of the extension's name (without the leading ``X'' which would normally signal that it is a non-standard extension) followed by a space (U+0020 in Unicode) and the actual data of the message. Non-standard extensions are required to use this message type.
\end{description}

A server always responds to these messages with a message with one of the following types:

\begin{description}
\item [O] Message signals that no error occurred and may also contain the data the client requested.
\item [Z] Temporary failure. The client should retry their request later.
\item [D] Permanent failure.
\end{description}

In a case of a failure the message should include a human-readable description right after the first byte.

More specifically regarding the positive responses, a response to the \textbf{H} message is either empty or contains a list of extensions supported by the server delimited with a space (U+0020). An extension's name may be at most 31 bytes long and must not contain a space (U+0020), otherwise any UTF-8 valid byte sequence is allowed, although usage of uppercase ASCII letters and digits is recommended. Proprietary extensions' names must begin with ASCII character ``X''. Empty extension names are not allowed; this also includes an extension name ``X'', which is therefore treated as empty.

A positive response to the \textbf{Q} message contains the data corresponding to the given cryptoaddress within the specified service.

\subsection{Example of a Query}
In the following example, ``C $>$: '' denotes a message sent by a client and ``$<$ S: '' denotes a reply sent by a server. Only the content of cryptographic boxes is shown. The client requests a bitcoin address for a cryptoaddress \textit{johndoe@example.com}.

\begin{verbatim}
C >: 1:H,
< S: 1:O,
C >: 92:Q19:johndoe@example.com,64:000000000019d6689c085ae165831e934ff763ae46a2
     a6c172b3f1b60a8ce26f,,
< S: 35:O1NS17iag9jJgTHD1VXjvLCEnZuQ3rJDE9L,
\end{verbatim}

\section{Security Considerations}
\label{sec:security_considerations}
Incorporation of several components potentially opens more possibilities for attacks. We are therefore discussing security impact of each component.

On network level, a connection to and communication with a cryptoaddress server is secured by authenticated encryption. The server's public key is known to the client before initiation of the connection. Therefore, the communication is safe from intercepting and tampering attacks. However, before a client connects to a cryptoaddress server, they queries a DNS server in order to obtain a public key and an IP address of the cryptoaddress server. Unless secured, the DNS traffic is unencrypted and forgeable. Therefore, we put as a requirement to use DNSSEC to sign records leading to the SRV records and associated A/AAAA records, including these. Additionally, we recommend usage of DNSCurve to provide users with confidentiality of their DNS queries \cite{dnscurve}.

Unless a cryptoaddress server is operated by the same person or company who uses it for their cryptoaddresses, a certain amount of trust in the server's operator is needed. The risk is that the operator could replace user's cryptographic identifier with their own one or one of some colluding party. For static cryptoaddress mappings a monitoring can be setup to detect potential dishonesty of the operator. However, there is no general solution for this issue yet.

Special considerations are needed for cryptoaddresses themselves. Threats related to them can be generalised to spoofing.

\subsection{Cryptoaddress Spoofing}
An attacker's goal we mainly consider here is mimicking cryptoaddresses in order to gain ability to forge cryptographic identifiers connected to the cryptoaddresses without the need to actually attack the cryptoaddress server.

Primary problem are visually similar characters, e.\,g., ``1'' (digit one) and ``l'' (lowercase L) in ASCII or ``c'' (Latin lowercase C; U+0063) and ``c'' (Cyrillic lowercase es; U+0441) in Unicode. There is no straightforward solution to this issue. Indeed, a cryptoaddress formed by the alternative characters (e.\,g., ``john@examp1e.org'' with lowercase L replaced by digit one) might be the authentic cryptoaddress and not a forgery.

Local part of a cryptoaddress may be further affected by the permission to include any valid UTF-8 character. Unicode enables some characters to be either formed by a single Unicode character or by a combination of two or more Unicode characters \cite{unicode-sec}. An example of such a character is the German ``ä'': this is either a single character U+00E4 or a combination of ``a'' and the combining diacritical mark umlaut---U+0061 and U+0308.

Furthermore, special or unprintable characters could be inserted into an otherwise legitimate cryptoaddress. Client applications should not interpret these characters and instead signalise their presence in the cryptoaddress. This is already a common practice with today's GUI applications.

Note that the risk of spoofing of cryptoaddresses, e.\,g., in the context of e-mails or web pages, is comparable with the risk of spoofing of cryptographic identifiers which a particular cryptoaddress represents, although spoofing of the original identifiers might be easier due to their random-looking nature \cite{mastering_bitcoin}.

\section{Future Work}
\label{sec:future_work}
In this paper we present the concept of cryptoaddresses in its basic form. Future work will focus on possible extensions, e.\,g., providing various information to authenticated users. In case of cryptocurrencies this may include information about index of the last issued dynamic address.

Another challenge, which is not yet covered, is the unsecured nature of information provided by a cryptoaddress server. An efficient method of preventing the server from being able to undetectably forge sensitive information, would reduce the level of trust in the server's operator that is currently needed.

\section{Conclusion}
\label{sec:conclusion}
Many systems whose central features include usage of cryptography to secure their functioning present users with long random-looking identifiers. These identifiers are supposed to be shared and to identify users or their resources within the systems. To improve usability of such systems, we propose in this paper a decentralised system of cryptoaddresses which resemble well-known e-mail addresses. The system enables cryptoaddresses to be mapped back to the original identifiers.

Our proposed system includes leveraging of the existing DNS system secured by its DNSSEC extension; implicit distribution of public keys of servers; and a design of a simple communication protocol between a cryptoaddress client and a server, based on top of a secure communication channel. Although primarily aimed at cryptocurrencies, the presented system is flexible, extensible and can be used for a variety of systems with cryptographic identifiers.

\bibliographystyle{eptcs}
\bibliography{biblio.bib}
\end{document}